\def\Journal#1#2#3#4{{#1}\, {\bf #2}, #3 (#4)}
  
\def\PRL{Phys.\ Rev.\ Lett.}

\documentclass[
    ,final            
  ]
  {aipproc}

\layoutstyle{6x9}
\usepackage{epstopdf}

\begin{document}

\title{Charged Particle Production at High Rapidity in p+p Collisions at RHIC}


\author{R. Debbe for the BRAHMS Collaboration}{
  address={Brookhaven National Laboratory}
}

\begin{abstract}
 This report describes the recent analysis of identified charged particle production at high rapidity 
performed on data collected from p+p collisions at RHIC ($\sqrt{s}=200 GeV$). The extracted invariant 
cross-sections compare well to  NLO pQCD calculations. However, a puzzling high yield of protons at high rapidity and
$p_{T}$ has been found.

\end{abstract}

\maketitle


\section{Introduction}

The p+p system is used in many cases as a paradigm of the simple hadronic interaction. In particular, nuclear
effects in p(d)+A and A+A collisions are extracted through their suitably normalized comparison to p+p
collisions. In reality the p+p system is still waiting for a truly from-first-principle QCD based description
that will include the production of particles and the distribution of the beam baryon number after the collision.
As part of the RHIC heavy-ion and spin programs, p+p collisions at $\sqrt{s}=200$ GeV have been produced during several
runs providing a great opportunity to revisit the work done at SppS \cite{UA2rapidity, UA7rapidity} and especially 
to fill the gaps left in the study of particle production at high rapidity. This report describes the 
production of charged particles in p+p collisions measured by the BRAHMS Collaboration and their comparison to
Next-to-Leading-Order perturbative Quantum Chromodynamics (NLO pQCD) calculations. It also presents a 
puzzling high yield of protons at high rapidity and transverse momentum.   

\section{The BRAHMS setup and analysis details.}

The results reported here were obtained with the BRAHMS Forward Spectrometer (FS) that consists of five 
tracking stations measuring the momentum of charged particles deflected in the field of three warm magnets. 
The particle identification for this work was performed entirely with the BRAHMS Ring Imaging Cherenkov (RICH) 
detector.
The efficiency of this detector was estimated from the data and reaches a constant value
of 97\%.
The data used in this analysis were collected with two main triggers, a minimum bias trigger built from the
signal produced in several stations of Cherenkov radiators placed near the beam pipe on both sides of the 
nominal interaction point called CC detectors, and a spectrometer trigger used to select events with tracks in the FS spectrometer. The CC detectors have an 
angular coverage the extends from 
$\eta = 3.26$ to $\eta = 5.25$. More details of the BRAHMS experimental setup can be found in \cite{BRAHMSNIM}.
The CC detectors have been included into a thorough simulation of the 
BRAHMS setup,
and it is estimated that they cover $70 \pm 1 \%$ of the 41mb total inelastic p+p cross-section 
at $\sqrt{s}=200$ GeV.
The spectrometer trigger was built with signals 
from three scintillator hodoscopes mounted on the spectrometer. 
Simulated events generated with PYTHIA \cite{Pythia} were processed with
a GEANT simulation of the BRAHMS setup,  
the efficiency of the spectrometer trigger to detect truly minimum biased events was extracted from that 
simulation and was found to be equal to $82.6 \pm 0.8\%$ and fairly independent of $p_{T}$.

\begin{figure}
  \includegraphics[height=.4\textheight]{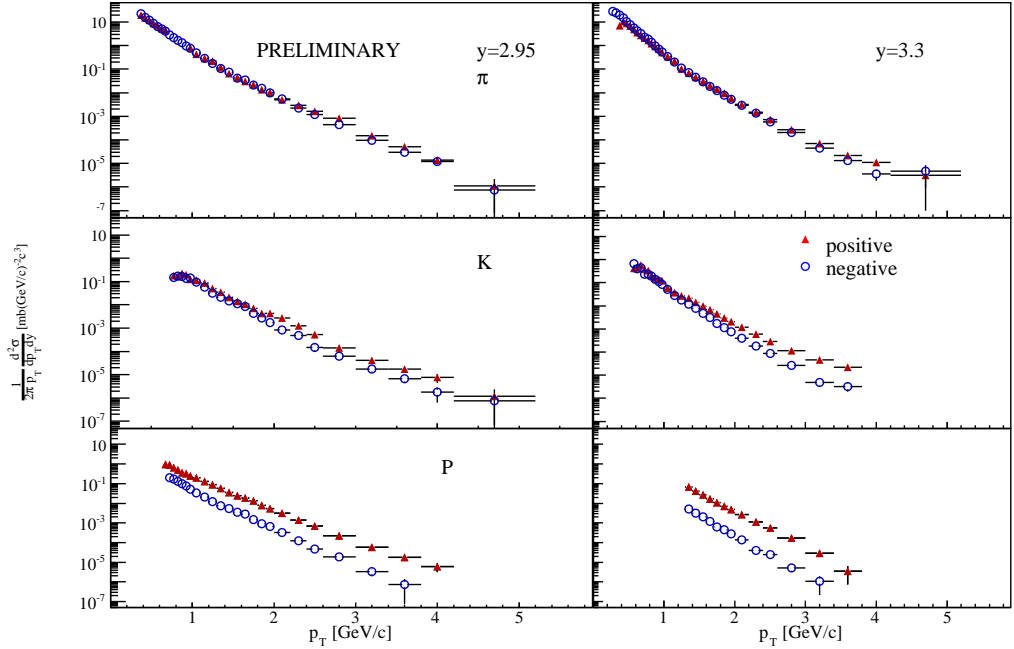}
  \caption{Invariant cross-section for production of pions, kaons and protons. See text for details.}
  \label{fig:spectra}
\end{figure}

Figure \ref{fig:spectra} shows the invariant cross-section for pions (top), kaons (middle), and protons (bottom) at two
rapidities y=2.95 on the left and y=3.3 on the right.
In each panel the positive particles are shown with filled
triangles and the negative ones with open circles. The spectra were extracted from narrow rapidity bins 
($\Delta y = 0.1$) centered in regions of the spectrometer acceptance with the highest reach in $p_{T}$ and may 
include up to five magnetic field settings.
The spectra have been corrected for the spectrometer acceptance, tracking reconstruction inefficiencies that 
range from 80-90\%, decays in flight, absorptions in the spectrometer and trigger bias.

\section{Comparisons with NLO pQCD calculations}

The differential cross-sections at y=2.95 have been compared to NLO pQCD calculations  \cite{Vogelsang} that use a 
modified 
version of the "Kniehl-Kramer-Potter" (KKP) set of fragmentation functions \cite{KKP} referred here as mKKP as 
well as the "Kretzer" set \cite{Kretzer}. Both calculations have been done with equal values of renormalization 
and factorization scales $\mu = p_{T}$. 
The functions from the KKP set fragment into sums: $\pi^{+}+\pi^{-}$, $K^{+}+K^{-}$ and $p+\bar{p}$. 
 Modifications were necessary  to obtain fragmentation into separate charges for $\pi $ and $K$:
The fragmentation function for $u$ and $\bar{d}$ quarks into $\pi^{+}$ were obtained as the product 
$D_{u}^{\pi^{+}}=(1+z)D_{u}^{\pi^{0}}$ with $D_{u}^{\pi^{0}}=\frac{1}{2}D_{u}^{\pi^{+}+\pi^{-}}$ and the functions 
fragmenting $\bar{u}$ and $d$ quarks into $\pi^{+}$ are obtained in similar fashion but the factor changes to $(1-z)$.
Similar operations are performed for fragmentation functions producing negative pions. In the case of kaons the operations
involve the $u, \bar{u}, s, \bar{s}$ quarks.

\begin{figure}
  \includegraphics[height=.3\textheight]{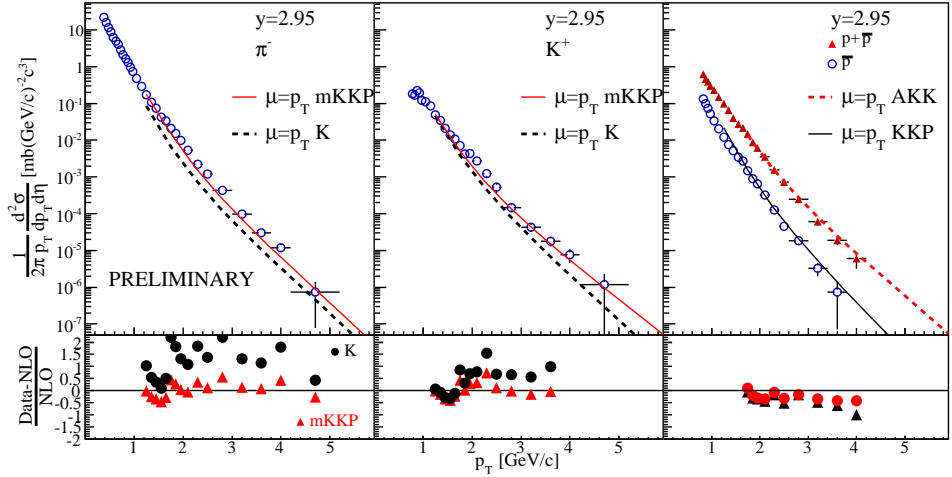}
  \caption{Comparison between the measured cross sections for the production of negative pion, positive kaons 
and $p+\bar{p}$  at y=2.95 and the NLO calculations at the same rapidity. Calculations including the mKKP set 
of fragmentation functions are shown with a full line (red online), the ones using the K set are shown
with dashed line (blue in online). The $p+\bar{p}$ sum is shown with filled triangles in the right panel 
and is well described by the calculations including the 
AKK set (shown with a dashed curve (red online). The same panel displays the anti-proton cross-section
with open circles and the NLO calculation for $p+\bar{p}$ using the KKP set of functions (full curve). 
The ratios (Data-NLO)/NLO 
for each of the comparisons mentioned above is shown in the three bottom panels. Filled triangles (red online)
 are used for the mKKP set and filled circles for the K set for pions and kaons. The comparison of the AKK and
the measured $p+\bar{p}$ cross-section is shown with filled circles (red online) and the one for the KKP
set and anti-protons is shown with filled triangles.}
  \label{fig:NLO}
\end{figure}

\begin{figure}
  \includegraphics[height=.3\textheight]{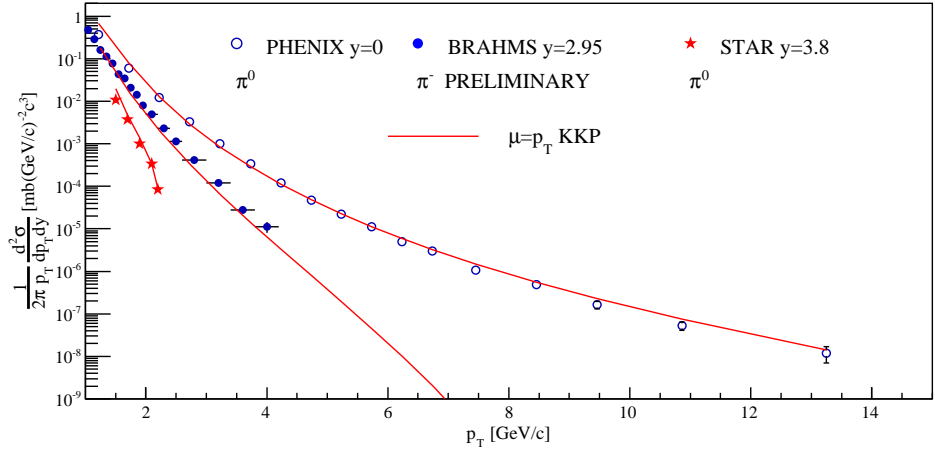}
  \caption{Invariant cross-sections as function of $p_{T}$ measured in p+p collisions at RHIC 
$\sqrt{s} = 200$ GeV by PHENIX at y=0 (open circles), BRAHMS at y=2.95 (filled circles, blue online)  and STAR at y=4 
(stars, red online). 
The curve is the NLO calculation with the KKP set of fragmentation functions. The errors are statistical. }
  \label{fig:RHIC_NLO} 
\end{figure}

The agreement between the measured cross-sections for negative pions and positive kaons at y=2.95 shown in the left 
and middle panels of Fig. \ref{fig:NLO} is remarkable. The calculation that incorporates the mKKP set is the one 
that describes best the produced particle distributions. Similar agreements have also
been seen at mid-rapidity for $\pi^{0}$ measured by PHENIX \cite{PHENIXpi0}, and at high rapidity,  
for the STAR $\pi^{0}$ measurement \cite{STARppForward}. Figure \ref{fig:RHIC_NLO} displays all three measurements showing
how the distributions get stepper at higher rapidities while the NLO calculations using the KKP set reproduce the 
data within four units of rapidity starting at surprisingly low values of $p_{T}$.   

\section{Ratios}

\begin{figure}
  \includegraphics[height=.3\textheight]{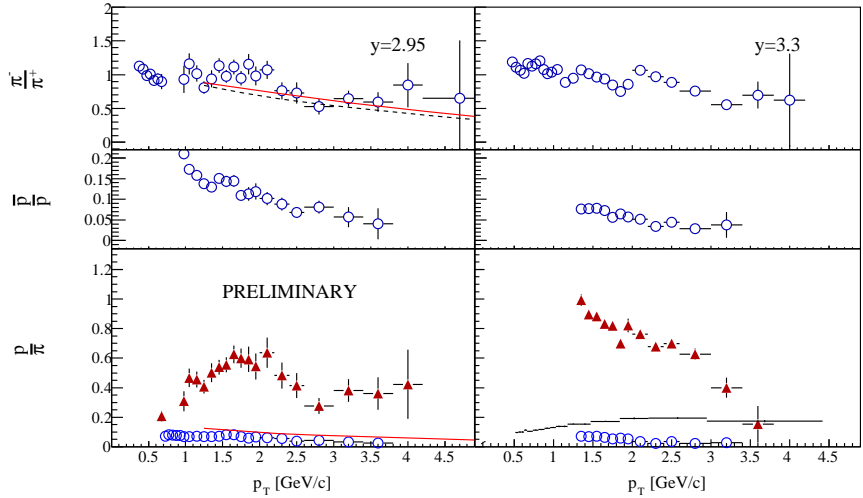}
  \caption{The $\pi^{-}/\pi^{+}$ ratios at both rapidities are shown on the top panels, the $\bar{p}/p$ ratio
is shown in the middle panel, the bottom panels show the $p/\pi^{+}$ and $\bar{p}/\pi^{-}$ ratios.
Data from $e^{+}e^{-}$ annihilations is shown with a dashed line \cite{epluseminus}. The smooth curves are obtained
from the NLO calculations of Fig. \ref{fig:NLO}.}
  \label{fig:ratios}
\end{figure}

Figure \ref{fig:ratios} shows the ratio of anti-particle to particle at both rapidities, as well as the ratio of
protons to pion. The results obtained at y=2.95 are shown on the left column and the ones at y=3.3 on the right 
one. The ratio $\pi^{-}/\pi^{+}$ shown on the top panels tends toward 1/2 in agreement with 
a simple valence quark counting in the proton beam. This hints to a possible dominance of valence quark
fragmentation in the pion productions at these high rapidities. The ratio of $\bar{p}/p$ shown in the two middle 
panels is small and drops
fast as the $p_{T}$ increases, indicating again the importance of beam protons. How did these protons
acquire their moderately high $p_{T}$ is an unknown. The successful reproduction of STAR proton 
cross-section at mid-rapidity in p+p collisions \cite{STARprotons} with NLO calculations that make use of a 
newly released set of fragmentation functions dubbed the AKK set \cite{AKK}, has been advanced as an indication of the
important contribution of the gluon fragmentation into protons or anti-protons. The same comparison,
done with the $p+\bar{p}$ sum at y=2.95, can be seen in the right-most panel of Fig. \ref{fig:NLO}.
The calculation lies almost on top of the data points but it is in a clear contradiction with the measured 
 $\bar{p}/p$ ratios shown in the middle panels of Fig. \ref{fig:ratios}. Gluons should fragment into 
$p \bar{p}$ pairs and will produce $\bar{p}/p$ with values close to 1 similar to the ones measured at mid-rapidity. 
The fact that the value
of the  $\bar{p}/p$ ratio at high rapidity is so small invalidates the use of the AKK set at 
high rapidity. Figure \ref{fig:NLO} also shows the NLO calculation with the KKP set for the sum $p+\bar{p}$ in
good agreement with the measured anti-protons cross-section. This could be a hint that factorized QCD and the 
parametrization of the fragmentation functions can describe the physics of produced particles but 
another approach is needed in order to include the protons and reach a truly complete description of high energy p+p collisions.   

The two bottom panels of Fig. \ref{fig:ratios} show the ratios $p/\pi^{+}$ (filled triangles, red online)
 and $\bar{p}/\pi^{-}$ (open circles) as function of 
$p_{T}$ and at rapidity y=2.95 (left) and y=3.3 (right). The difference between the ratios of negative particles
(produced) and positives is remarkable and shows that a factorized pQCD based description fails for protons. The
ratio $\bar{p}/\pi^{-}$ is closer to a rough parametrization extracted from $e^{+} e^{-}$ annihilations. 

\section{Summary}

We have presented the recently measured cross-sections for pions, kaons and protons at high rapidity in p+p collisions at RHIC. 
The NLO pQCD calculations incorporating the KKP fragmentation functions describe the pion and kaon cross-sections.
Protons at high rapidity and $p_{T}$ appear in unusually high numbers that do not fit into a factorized pQCD
description based on parton-parton interactions.


We thank Werner Vogelsang for providing us with the NLO pQCD calculations shown in this report.
This work was supported by 
the Office of Nuclear Physics of the U.S. Department of Energy, 
the Danish Natural Science Research Council, 
the Research Council of Norway, 
the Polish State Committee for Scientific Research (KBN) 
and the Romanian Ministry of Research.

\end{document}